\documentclass{llncs}
\usepackage[latin1]{inputenc}
\usepackage[english]{babel}
\usepackage{amsmath}
\usepackage{amsfonts}
\usepackage{amssymb}
\usepackage{graphicx}
\usepackage{subfigure}
\usepackage{tikz}
\usepackage{tkz-graph}
\usetikzlibrary{shapes,arrows}
\usepackage{textcomp}
\usepackage[customcolors]{hf-tikz}
\usetikzlibrary{arrows,positioning,shapes,snakes}
\usepackage{blkarray}
\usetikzlibrary{matrix}

\author{Charalampos Kyriakopoulos\inst{1} \and Pascal Giehr\inst{2} \and Alexander L\"uck\inst{1} \and J\"orn Walter\inst{2} \and Verena Wolf\inst{1}}
\institute{Department of Computer Science, Saarland University, Saarbr\"ucken, Germany \and Department of Biological Sciences, Saarland University, Saarbr\"ucken, Germany
}
\author{Charalampos Kyriakopoulos\inst{1} \and Pascal Giehr\inst{2} \and Alexander L\"uck\inst{1} \and 
J\"orn Walter\inst{2} \and Verena Wolf\inst{1}}
\institute{Department of Computer Science, Saarland University, Saarbr\"ucken, Germany \and Department of Biological Sciences, Saarland University, Saarbr\"ucken, Germany 
}

\title{
 A Hybrid HMM Approach for the Dynamics of DNA Methylation}
\begin{document}
\maketitle

%
%

\begin{abstract}
 The understanding of mechanisms that control epigenetic changes is an important research area in modern functional biology.
Epigenetic modifications such as DNA methylation are in general very stable over many cell divisions. 
DNA methylation can however be subject to specific and fast changes over a short time scale  even in non-dividing (i.e. not-replicating) cells. 
Such dynamic DNA methylation changes are caused by a combination of active demethylation and de novo methylation processes which have not been investigated in integrated models.
	
Here we present a hybrid (hidden) Markov model to describe the cycle of methylation and demethylation over (short) time scales. 
Our hybrid model decribes several molecular events either happening at deterministic points (i.e. describing mechanisms that occur only during cell division) and other events   occurring at random time points.
We test our model on mouse embryonic stem cells using time-resolved data. We predict   methylation changes and estimate the efficiencies of the different 
  modification steps related to DNA methylation and demethylation.
\end{abstract}

\section{Introduction}
 All cells of a multi-cellular organism share the same DNA sequence, yet, depending on location and cell type, display distinct cellular programs as a result of controlled gene expression.
Hence, the expression of genes is regulated by epigenetic factors such as DNA methylation.
In mammals, the methylation of DNA is restricted to the C5 position of cytosine (C) and mostly appears in a CpG di-nucleotide sequence \cite{ehrlich1982amount,feng2010conservation}.
The palindromic nature of CpG positions provides a symmetry which, after DNA replication, allows the stable inheritance of the methylation ``signal''. 
Methylation of C to 5-methyl cytosine (5mC) is catalysed by a certain enzyme family,
the DNA methyltransferases (Dnmts) \cite{bestor1983two,okano1998cloning}.
Three conserved and catalytic active family members are associated with the methylation of DNA.
Dnmt1 is mainly responsible for maintenance methylation after DNA replication,  
i.e. the enzyme mainly reestablishes the methylation pattern on the newly synthesized daughter strand \cite{hermann2004dnmt1}  according to the inherited information of the parental DNA strand.
The enzymes Dnmt3a and Dnmt3b perform
\emph{de novo} methylation, where new methyl groups are added on unmethylated Cs \cite{okano1999dna}.
However, there is evidence that this separation of tasks is not definite and that all Dnmts may carry out all tasks to a certain extent \cite{arand2012vivo}.

Once established, 5mC can be further modified by oxidation to 5-hydroxyme\-thyl cytosine (5hmC), which  can again be oxidized to 5-formyl cytosine (5fC). Then, 5fC can eventually be converted to 5-carboxy cytosine (5caC).
All these processes are carried out by 
the ten-eleven translocation (Tet) enzymes \cite{ito2011tet,tahiliani2009conversion}.
A considerable level of 5hmC can be found in many cells types and its occurrence has been connected to gene regulation as well as genome wide loss of DNA methylation \cite{globisch2010tissue,kriaucionis2009nuclear}.
In contrast, 5fC and 5caC are far less abundant and their particular functions remain more illusive.
Nevertheless, studies suggest that both oxidized cytosine variants  function as intermediates during enzymatic removal of 5mC from the DNA \cite{globisch2010tissue}. 

In general, DNA methylation can be removed in two ways.
First, after DNA replication, the absence or blocking of maintenance methylation will result in a passive DNA methylation loss with each cell division  (\emph{passive demethylation}).
Second, generated 5fC or 5caC is enzymatically removed from the DNA and subsequently replaced by unmodified cytosine (\emph{active demethylation})  \cite{cardoso1999dna,hashimoto2012recognition,he2011tet,maiti2011thymine}.

While 
 DNA replication  and the associated maintenance methylation happens only once per cell division cycle, \emph{de novo} methylation and the modification processes of 5mC via oxidation, as well as active demethylation, may happen at arbitrary time points. 
With purely discrete hidden Markov models (HMM), as used in \cite{arand2012vivo} and   \cite{Giehr2016}, 
it is difficult to describe multiple instances of \emph{de novo} methylation or other modification events for a given CpG during one cell division cycle.
Here, we introduce a hybrid HMM
to describe  the dynamics of active demethylation.
 It distinguishes between events at fixed time points (cell division and maintenance methylation) and events at random time points (\emph{de novo} methylation, oxidations, active demethylation).
By applying our model to a data set of a single copy gene from mouse embryonic stem cells, we were able to accurately predict the frequency of the observable CpG states and the levels of the hidden states, which correspond to the different modified forms of C.
Furthermore, we show how to estimate the enzymatic
reaction efficiencies using a maximum likelihood approach.

Compared to previous models for describing DNA methylation, 
 we here propose an   approach that takes into account both, the rather fixed and deterministic timing of cell division and the random nature of the enzymatic processes.
In this way, we are able to present a   model that realistically describes the dynamics of DNA methylation and improves our understanding of  active demethylation.

The paper is organized as follows: 
In Section 2 we give the necessary biological and mathematical background, i.e. we explain   passive and active demethylation in more detail, describe the model and explain the parameter estimation procedure.
In Section 3 the results are discussed and in Section 4 we conclude our findings. 

\section{Model}
\subsection{Passive and active demethylation}
One can distinguish between two different ways of losing methylation at cytosines:
After 
cell division a new (daughter) strand of the DNA is synthesized.
Initially, all cytosines of the daughter strand 
are 
unmethylated, while the methylation states of cytosines on the parental strand remain unchanged.
Maintenance methylation, which happens during the replication process, is used to re-establish the methylation
pattern at the newly synthesized strand. However, 
the absence or inhibition of maintenance methylation causes a loss of 5mC with each replication step.
This DNA replication dependent loss has been termed \emph{passive demethylation}. 

\begin{figure}[t]
\begin{center}
		\begin{tikzpicture}[minimum size=1cm,scale=0.5,->,>=stealth',thick] 	

		\node (4) at (34, 3.5) [fill= yellow, circle,draw] {$h$};
		\node (5) at (30,0) [fill= red, circle,draw] {$m$};
		\node (6) at (38,0) [fill=cyan!40, circle,draw] {$f$};
		\node(7) at (34,-3.5) [fill = blue, circle,draw] {$u$};

	\tikzset{LabelStyle/.style={sloped, above,yshift = 0cm, xshift=.0cm}}
	\Edge[label={$\eta$}, style={bend left=30}](5)(4) 
	\Edge[label={$\phi$}, style={bend left=30}](4)(6)
	\tikzset{LabelStyle/.style={sloped, below,yshift = -0cm, xshift=.0cm}}
	\Edge[label={$\delta$}, style={bend left=30}](6)(7)
	\Edge[label={$\mu_d$}, style={bend left=30}](7)(5)

	\end{tikzpicture}
\end{center}
\caption{Schematic representation of \emph{de novo} methylation and the active demethylation loop, where we use the following notation for the methylation states: (Unmethylated) C is   denoted by $u$, 5mC by $m$, 5hmC by $h$, and 5fC or 5caC by $f$. 
The corresponding enzymatic reaction rates are $\mu_d$ (\emph{de novo} methylation), $\eta$ (oxidation), $\phi$ 
(formylation), and $\delta$ (demethylation).}
\label{Fig:MethCycle}
\end{figure}
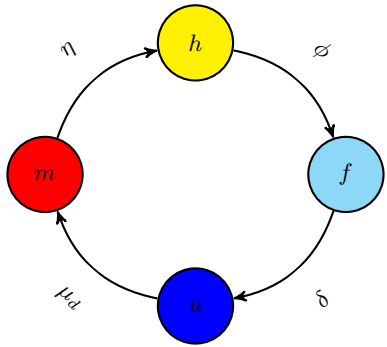

For \emph{active demethylation} it is assumed that   oxidation of 5mC to 5fC or 5caC via 5hmC and a subsequent enzymatic removal of the oxidative cytosine from the DNA occurs (cf. Fig \ref{Fig:MethCycle}).
Since \emph{de novo} methylation as well as \emph{active demethylation} are replication independent, the loop depicted in Fig. \ref{Fig:MethCycle} 
can be traversed   multiple times within one cell cycle. 
In measurements, we see Cs in all stages, i.e., either methylated, oxidized (5hmC/5fC/5caC) or unmodified.

\subsection{Hybrid Markov model}
In this section, we present a model that describes the state changes of a  single CpG over time. It can be seen as a hybrid extension of
previous discrete-time models \cite{arand2012vivo,Giehr2016}.
The (hidden) states of a CpG correspond to the set of all pairs of the four possibilities in Fig.   \ref{Fig:MethCycle}, i.e., $\{u,m,h,f\}^2$, because it contains a C on both strands of the DNA.
We split the transitions of our model into transitions that occur at  fixed times and those that occur at random times.
This results in a mixture of a discrete time Markov chain (DTMC) and a continuous time Markov chain (CTMC).
In the following we will refer to the events or transitions that occur at   fixed time points as \emph{discrete part} of the model, while we refer to the other events or transitions at the random time points as \emph{continuous part} of the model. 
 
We assume that  cells divide after  a fixed time interval (usually every 24 hours). Hence, these events correspond to deterministic transitions at fixed times.
During   cell division one strand is kept as it is (parental strand) and all methylation states and its modifications remain unchanged, while one DNA strand is newly synthesized (daughter strand) and therefore contains only unmethylated cytosine.
Consequently, after cell divison a CpG that was modified on both sides becomes a CpG that is unmodified on one side.  
Since the parental strand is chosen at random, the probability for each of the two successor states (corresponding to the state of the CpG in the two daughter cells)   is 0.5.
The full transition probability matrix $\mathbf{D}$ for  cell division  is shown in Tab.~\ref{Tab:D}. Note that the cell division matrix is time independent.

\renewcommand{\arraystretch}{1.25}
\setlength{\tabcolsep}{3pt} 
\begin{table}[tb]
\caption{Cell division matrix $\mathbf{D}$.}
\label{Tab:D}
\begin{center}
\begin{tabular}{c|cccccccccccccccc}
~ & $uu$& $um$ & $uh$ & $uf$ & $mu$ & $mm$ & $mh$ & $mf$ & $hu$ & $hm$ & $hh$ & $hf$ & $fu$ & $fm$ & $fh$ & $ff$ \\\hline
$uu$ & 1 & 0 & 0 & 0 & 0 & 0 & 0 & 0 & 0 & 0 & 0 & 0 & 0 & 0 & 0 & 0\\
$um$ & 1/2 & 1/2 & 0 & 0 & 0 & 0 & 0 & 0 & 0 & 0 & 0 & 0 & 0 & 0 & 0 & 0\\
$uh$ & 1/2 & 0 & 1/2 & 0 & 0 & 0 & 0 & 0 & 0 & 0 & 0 & 0 & 0 & 0 & 0 & 0\\
$uf$ & 1/2 & 0 & 0 & 1/2 & 0 & 0 & 0 & 0 & 0 & 0 & 0 & 0 & 0 & 0 & 0 & 0\\
$mu$ & 1/2 & 0 & 0 & 0 & 1/2 & 0 & 0 & 0 & 0 & 0 & 0 & 0 & 0 & 0 & 0 & 0\\
$mm$ & 0 & 1/2 & 0 & 0 & 1/2 & 0 & 0 & 0 & 0 & 0 & 0 & 0 & 0 & 0 & 0 & 0\\
$mh$ & 0 & 0 & 1/2 & 0 & 1/2 & 0 & 0 & 0 & 0 & 0 & 0 & 0 & 0 & 0 & 0 & 0\\
$mf$ & 0 & 0 & 0 & 1/2 & 1/2 & 0 & 0 & 0 & 0 & 0 & 0 & 0 & 0 & 0 & 0 & 0\\
$hu$ & 1/2 & 0 & 0 & 0 & 0 & 0 & 0 & 0 & 1/2 & 0 & 0 & 0 & 0 & 0 & 0 & 0\\
$hm$ & 0 & 1/2 & 0 & 0 & 0 & 0 & 0 & 0 & 1/2 & 0 & 0 & 0 & 0 & 0 & 0 & 0\\
$hh$ & 0 & 0 & 1/2 & 0 & 0 & 0 & 0 & 0 & 1/2 & 0 & 0 & 0 & 0 & 0 & 0 & 0\\
$hf$ & 0 & 0 & 0 & 1/2 & 0 & 0 & 0 & 0 & 1/2 & 0 & 0 & 0 & 0 & 0 & 0 & 0\\
$fu$ & 1/2 & 0 & 0 & 0 & 0 & 0 & 0 & 0 & 0 & 0 & 0 & 0 & 1/2 & 0 & 0 & 0\\
$fm$ & 0 & 1/2 & 0 & 0 & 0 & 0 & 0 & 0 & 0 & 0 & 0 & 0 & 1/2 & 0 & 0 & 0\\
$fh$ & 0 & 0 & 1/2 & 0 & 0 & 0 & 0 & 0 & 0 & 0 & 0 & 0 & 1/2 & 0 & 0 & 0\\
$ff$ & 0 & 0 & 0 & 1/2 & 0 & 0 & 0 & 0 & 0 & 0 & 0 & 0 & 1/2 & 0 & 0 & 0\\
\end{tabular}
\end{center}
\end{table}

\begin{figure}[tb]
\begin{center}
\includegraphics[scale=0.35]{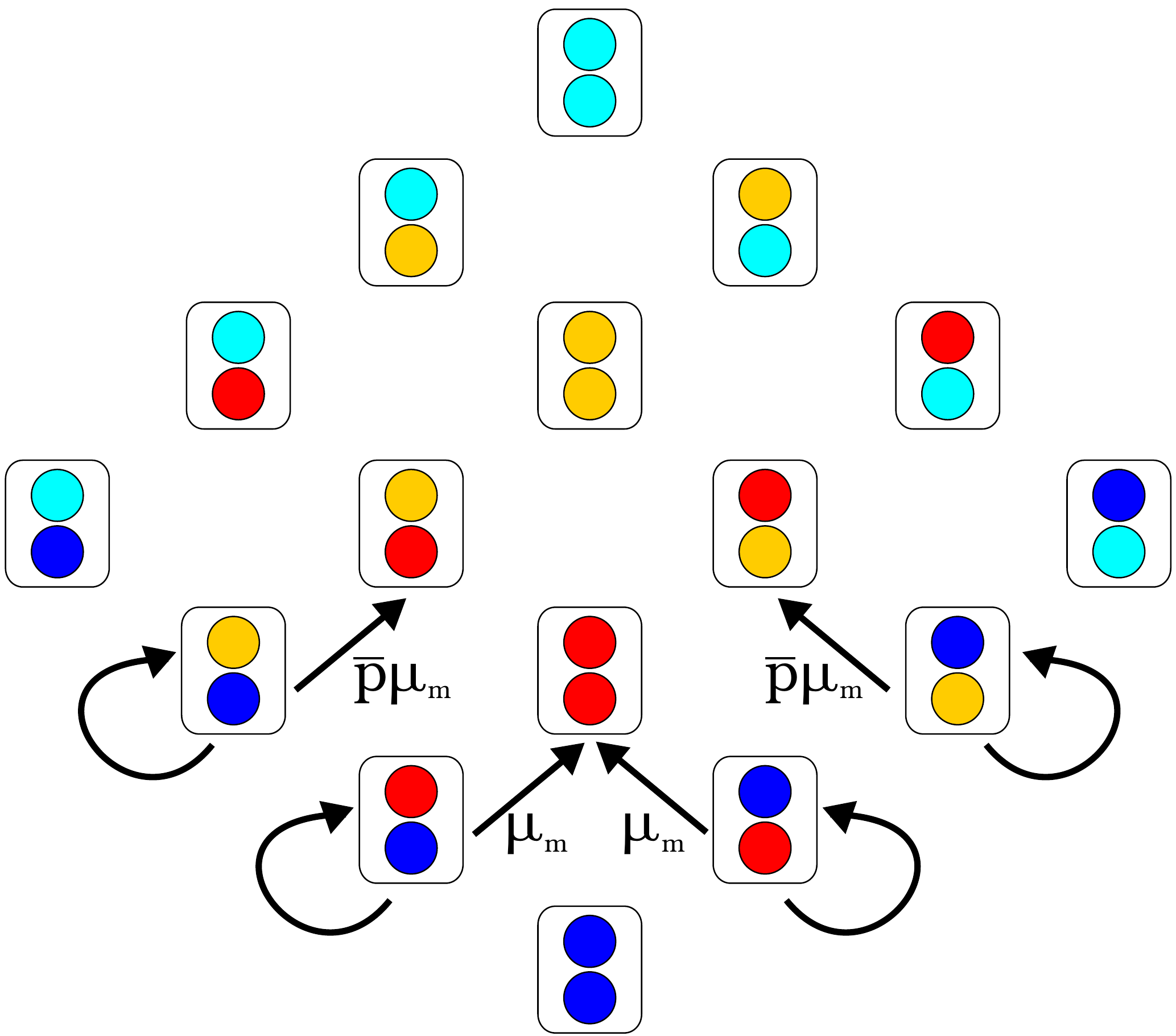}
\end{center}
\caption{Maintenance methylation events 
occur at (fixed) times $t_1,t_2,\ldots,t_n$ and belong therefore to the disrete part of the model.
 Each state is represented by two colored dots, one for each C on the two strands of the DNA. Unmethylated C is blue, 5mC red, 5hmC yellow and 5fmC cyan. The arrows indicate the possible transitions. Note that we omitted the self loops with probability $1$ for states where no transition is possible. The four shown self loops have probability $1-\mu_m$ or $1-\bar{p}\mu_m$ respectively.}
\label{Fig:M}
\end{figure}

Maintenance methylation, i.e. methylation events that 
occur on hemimethylated CpGs to reestablish methylation patterns, is known to be linked to the replication fork \cite{leonhardt1992targeting}.
We therefore consider   maintenance to occur together with the cell division at the same fixed time points. 
Hence, cell division and maintenance can be described by a (discrete-time) Markov chain whose transition probability matrix $\mathbf{P}(t)$ is defined in the sequel in Eq.~\eqref{eq:P}.
Maintenance may happen at the daughter strand if there is a methylated C on the parental strand, i.e. on hemimethylated CpGs ($um$ or $mu$), with probability $\mu_m(t)$.
Since it is reasonable to assume that hemihydroxylated CpGs ($uh$ or $hu$) have different properties compared to hemimethylated CpGs in terms of maintaining existing methylation patterns, we describe the maintenance probability for hemihydroxylated CpGs as follows.
Let $p$ be the probability that 5hmC is recognized as unmethylated by maintenance enzymes, i.e., the enzyme will not perform maintenance of a hemihydroxylated CpG.
Then, the maintenance probability of such a CpG  is  given by $\bar{p}\mu_m(t)$, where $\bar{p}=1-p$.
Note that there is no equivalent probability for $uf$ or $fu$, i.e. we assume that  CpGs with 5fC or 5caC at one strand are not maintained \cite{ji2014effects}.
 The transition probability matrix for maintenance events
 is illustrated in Fig.~\ref{Fig:M}, where we omitted the time dependency of $\mu_m(t)$.
Whenever no transition is possible, i.e. there is only a self loop for this state (omitted in Fig.~\ref{Fig:M}) the corresponding diagonal entry in the matrix is 1.
Where a transition is possible we set the corresponding (off-diagonal) entry in the matrix to its transition probability and the diagonal entry to 1 minus its transition probability.
All other entries in the matrix are 0.
One discrete step of the corresponding DTMC corresponds to one cell division, including maintenance methylation.
Hence, its transition probability matrix is defined as
\begin{equation}
\mathbf{P}(t)=\mathbf{D}\cdot\mathbf{M}(t).
\label{eq:P}
\end{equation}

\begin{figure}[tb]
\begin{center}
\includegraphics[scale=0.45]{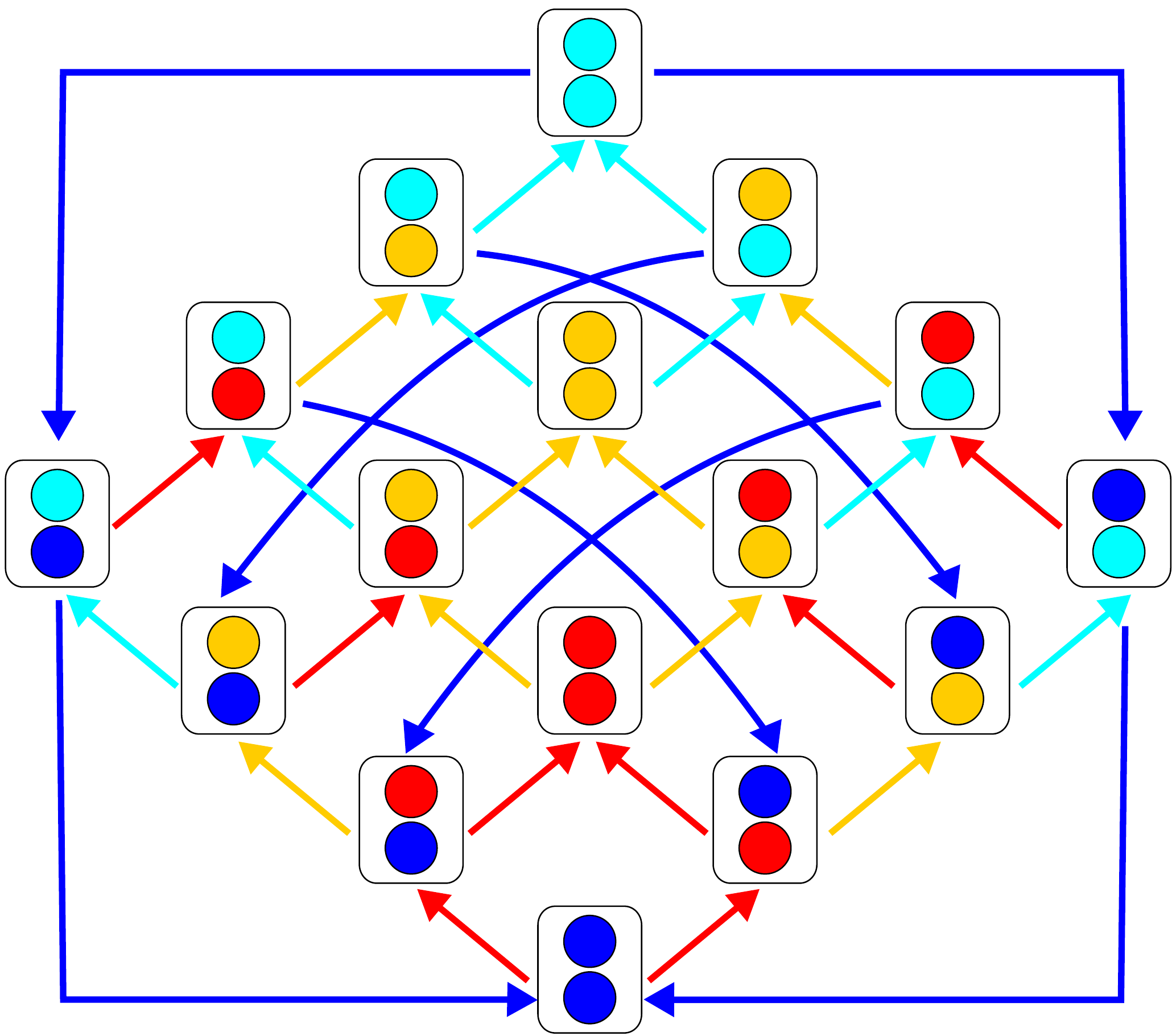}
\end{center}
\caption{Possible transitions in the continuous part of the model. Each state is represented by two colored dots, one for each C on the two strands of the DNA. Unmethylated C is blue, 5mC red, 5hmC yellow and 5fmC cyan. The arrows indicate the possible transitions, whereupon the colors indicate the different reactions and their rates, i.e. methylation with rate $\mu_d$ (red), oxidation with rate $\eta$ (yellow), formylation with rate $\phi$ (cyan) and active demethylation with rate $\delta$ (blue).}
\label{Fig:Q}
\end{figure}

Every other event may occur an arbitrary (unknown) number of times between two cell divisions at random time points and will 
be described by  a continuous-time Markov jump process.
These events are \emph{de novo} methylation ($u\rightarrow m$) with rate $\mu_d(t)$, hydroxylation ($m\rightarrow h$) with rate $\eta(t)$, 
formylation ($h\rightarrow f$) with rate $\phi(t)$ and active demethylation ($f\rightarrow u$) with rate $\delta(t)$ (cf. Fig~\ref{Fig:MethCycle}).
Note that all these events may happen on both strands, independent of the state on the complementary strand.
All possible reactions are shown in Fig.~\ref{Fig:Q}.
From these transitions the infinitesimal generator matrix $\mathbf{Q}(t)$ of the jump process can easily be inferred.
For the off-diagonal elements, we set the entries to the respective reaction rate if a reaction is possible between two states, indicated by a colored arrow in Fig.~\ref{Fig:Q}, and to $0$ if no reaction is possible.
The diagonal elements are then given by the negative sum of the off-diagonal elements of the respective row.

 In order to describe the time evolution let us first define the set of time points $T_d=\{t_1,t_2,\ldots,t_n\}$, at which     cell division and maintenance occur.
We assume that there are in total $n$ of these events.
At these time points $t_i$ the time evolution of the probability distribution of the states is given by
\begin{equation}
\pi(t_i+\Delta t)=\pi(t_i)\cdot\mathbf{P}(t_i),
\label{Eq:mult}
\end{equation}
where $\Delta t$ is the duration of cell division and maintenance.
After cell division and maintenance the other events take place at random time points until the time point for the next cell division is reached.
During this interval $[t_i+\Delta t,t_{i+1}]$ the time evolution for $\pi(t)$ is obtained by solving the differential equation
\begin{equation}
\frac{d}{dt}\pi(t)=\pi(t)\cdot\mathbf{Q}(t).
\label{Eq:int}
\end{equation}
Note that since cell division and maintenance methylation occur at a time interval that is much shorter  than the time between two cell divisions, we assume here that these events occur instantaneously, i.e. we let $\Delta t\to 0$, which leads to a jump of the distribution at $t_i$.
It holds that the left-hand and right-hand limit do not coincide (except for the special case of only $uu$ at time $t_i$), i.e.
\begin{equation}
\lim_{t\rightarrow t_i^{-}}\pi(t)=\pi(t_i^-)\neq\pi(t_i^-)\cdot\mathbf{P}(t_i^-)=\pi(t_i^+)=\lim_{t\rightarrow t_i^{+}}\pi(t).
\end{equation}
To resolve the ambiguity we set the value of $\pi$ at time $t_i$ to $\pi(t_i):=\lim_{t\rightarrow t_i^{+}}\pi(t)$, which means that we assume that at time $t_i$ the cell division and maintenance methylation have already happened.
Intuitively, we can obtain the solution for $\pi(t)$ numerically by alternating between multiplication with $\mathbf{P}$ (Eq.~\eqref{Eq:mult}) and integration of Eq.~\eqref{Eq:int}.
 
\subsection{Efficiencies}
As already discussed in \cite{Giehr2016} the methylation or modification rates may change over time, i.e. with the assumption of constant rates the behavior of the system can not be captured correctly.
We therefore introduce time dependent rates, which we call efficiencies.
The simplest way of making the efficiencies time dependent is to impose a linear form.
Let $r\in \{\mu_m,\mu_d,\eta,\phi,\delta\}$ be one of the reaction rates.
For each reaction we then define
\begin{equation}
r(t):=\alpha_r+\beta_r\cdot t,
\end{equation}
where $\alpha_r$ and $\beta_r$ are some new parameters to characterize the linear efficiency function.
To further compress the notation we write these new parameters into a vector $\vec{v}_r=(\alpha_r,\beta_r)$.

Note that in order to ensure identifiability of the parameters we have to introduce certain constrains.
Obviously since $\mu_m(t)$ and $p$ are probabilities, they have to be bound between zero and one, i.e. $0\leq\mu_m(t)\leq 1$ for all $t\in[0,t_{max}]$ and $0\leq p\leq 1$.
The other efficiencies have to be bound by some upper limit $ub$ since otherwise the methylation cycle may become arbitrarily fast and the optimization algorithm runs into identifiability problems.
We therefore require $0\leq r(t)\leq ub$ for all $t\in[0,t_{max}]$, where $r(t)\neq \mu_m(t)$.
Since typical average turnover times $\mathbb{E}[T_{turnover}]=\mu_d^{-1}+\eta^{-1}+\phi^{-1}+\delta^{-1}$ of the methylation cycle are in the order of $75$ to $120$ minutes in certain promotors of human cells \cite{kangaspeska2008transient,metivier2008cyclical}, a viable choice would be, for example, $ub=12$, i.e. each modification occurs on average not more frequent than 12 times per hour (not faster than every $5$ minutes).

\subsection{Conversion errors}
\begin{figure}[bt]
\begin{center}
\begin{tikzpicture}[scale=0.90, every node/.style={scale=0.90}]
  \matrix (m) [matrix of math nodes,row sep=3em,column sep=0.1em,minimum width=3em]
  {\text{\textbf{BS}} & ~& ~& ~& \text{\textbf{oxBS}} & ~& ~& ~& \text{\textbf{MAB-Seq}} & ~& ~& ~\\[-2.9em]
  ~& ~& ~& ~& [1em]& ~& ~& ~ & [1em]\text{C} & ~ & ~ &~ \\[-1.5em]
   \text{C} & 5\text{mC} & 5\text{hmC} & 5\text{fC}^* & \text{C} & 5\text{mC} & 5\text{hmC} & 5\text{fC}^* & \text{C} & 5\text{mC} & 5\text{hmC} & 5\text{fC}^* \\
     \text{U} & 5\text{mC} & 5\text{hmC} & 5\text{fU}^* & \text{U} & 5\text{mC} & 5\text{fU} & 5\text{fU}^* & \text{U} & 5\text{mC} & 5\text{hmC} & 5\text{fU}^* \\[-1.5em]
     \text{T} & \text{C} & \text{C} & \text{T} & \text{T} & \text{C} & \text{T} & \text{T} & \text{T} & \text{C} & \text{C} & \text{T} \\};
  \path[-stealth]
    (m-3-1) edge node [left] {$c$} (m-4-1)
    (m-3-2) edge node [left] {$d$} (m-4-2)
    (m-3-3) edge node [left] {$e$} (m-4-3)
    (m-3-4) edge node [left] {$g$} (m-4-4)
    (m-3-5) edge node [left] {$c$} (m-4-5)
    (m-3-6) edge node [left] {$d$} (m-4-6)
    (m-3-7) edge node [left] {$f$} (m-4-7)
    (m-3-8) edge node [left] {$g$} (m-4-8)
    (m-3-9) edge node [left] {$c$} (m-4-9)
    (m-3-10) edge node [left] {$d$} (m-4-10)
    (m-3-11) edge node [left] {$e$} (m-4-11)
    (m-3-12) edge node [left] {$g$} (m-4-12)
    
    (m-4-1) edge node [left] {} (m-5-1)
    (m-4-2) edge node [left] {} (m-5-2)
    (m-4-3) edge node [left] {} (m-5-3)
    (m-4-4) edge node [left] {} (m-5-4)
    (m-4-5) edge node [left] {} (m-5-5)
    (m-4-6) edge node [left] {} (m-5-6)
    (m-4-7) edge node [left] {} (m-5-7)
    (m-4-8) edge node [left] {} (m-5-8)
    (m-4-9) edge node [left] {} (m-5-9)
    (m-4-10) edge node [left] {} (m-5-10)
    (m-4-11) edge node [left] {} (m-5-11)
    (m-4-12) edge node [left] {} (m-5-12)
    
    (m-3-1) edge [red,dashed] node [right=0.25cm] {}  (m-4-2)
    (m-3-2) edge [red,dashed] node [right=0.25cm] {}  (m-4-1)
    (m-3-3) edge [red,dashed] node [right=0.25cm] {}  (m-4-1)
    (m-3-4) edge [red,dashed] node [right=0.25cm] {}  (m-4-2)
    (m-3-5) edge [red,dashed] node [right=0.25cm] {}  (m-4-6)
    (m-3-6) edge [red,dashed] node [right=0.25cm] {}  (m-4-5)
    (m-3-7) edge [red,dashed] node [right=0.25cm] {}  (m-4-6)
    (m-3-8) edge [red,dashed] node [right=0.25cm] {}  (m-4-6)
    (m-3-9) edge [red,dashed] node [right=0.25cm] {}  (m-4-10)
    (m-3-10) edge [red,dashed] node [right=0.25cm] {}  (m-4-9)
    (m-3-11) edge [red,dashed] node [right=0.25cm] {}  (m-4-9)
    (m-3-12) edge [red,dashed] node [right=0.25cm] {}  (m-4-10)
    
    (m-2-9) edge node [right] {$\mu$} (m-3-10)
    (m-2-9) edge [red,dashed] node [right=0.25cm] {}  (m-3-9);
\end{tikzpicture}
\end{center}
\caption{Cytosine conversions during chemical treatment and sequencing. 
The correct reactions with their respective probabilities are marked with black arrows 
, while the false reactions are shown with red dashed arrows. The probability for a false reaction is 1-``rate of correct reaction".}
\label{Fig:ConvErr}
\end{figure}
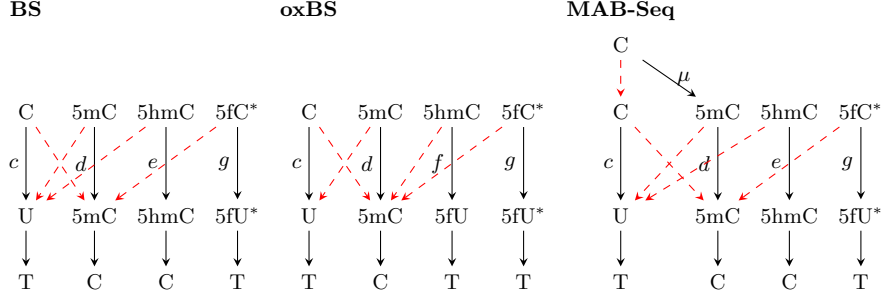

The actual state of a CpG can not be directly observed.
We therefore have to estimate the hidden states from sequencing experiments.
Since the different modifications of C might lead to the same observable states we perform three different kinds of sequencing experiments:
All sequencing strategies share a bisulfite treatment, which usually converts C and its oxidized variants 5fC and 5caC, summarized to 5fC$^*$, to uracil.
Additionally, in \emph{bisulfite sequencing} (BS) both 5mC and 5hmC remain unconverted 
while in {oxidative bisulfite sequencing} (oxBS) only 5mC is 
retained as C \cite{booth2012quantitative}.
A combination of both methods can therefore be used to estimate the amount of 5hmC.
In \emph{M. SssI assisted bisulfite sequencing} (MAB-Seq) 
at first all Cs in a CpG context are methylated and afterwards BS is applied.
Thus, if all conversions would happen without any errors, only 5fC$^*$ would be converted to T.
In order to capture the methylation pattern of CpG position at both complementary DNA strands, the distinct chemical treatments were combined with hairpin sequencing \cite{giehr2018two,giehr2018hairpin,laird2004hairpin}.
Regular reactions with their respective probabilities are marked with solid black arrows in Fig.~\ref{Fig:ConvErr}, while the possible false reactions are depicted with dashed red arrows.
Since every CpG consists of two Cs (one on each strand) with independent conversion errors, we get the conversion error for CpGs by multiplying the individual conversion errors.
A complete overview of all possible combinations for each of the three methods is shown in Tab.~\ref{Tab:ConvErr}.

\renewcommand{\arraystretch}{1}
\setlength{\tabcolsep}{0.5pt} 
\begin{table}
\begin{center}
\caption{Conversion errors for CpGs, where the rates for single cytosines are defined in Fig.~\ref{Fig:ConvErr}. We define $\bar{x}:=1-x$. Note that for MAB-Seq we also define $j:=\mu d+(1-\mu)(1-c)$.}
\label{Tab:ConvErr}
\begin{tabular}{c|cccc|cccc|cccc}
~ & \multicolumn{4}{|c}{bisulfite seq. (BS)} & \multicolumn{4}{|c}{ox. bisulfite seq. (oxBS)} & \multicolumn{4}{|c}{MAB-Seq}\\
~ & ~~TT~~ & ~~TC~~ & ~~CT~~ & ~~CC~~ & ~~TT~~ & ~~TC~~ & ~~CT~~ & ~~CC~~ & ~~TT~~ & ~~TC~~ & ~~CT~~ & ~~CC~~ \\\hline 
$~~uu~~$ & $c^2$ & $c\!\cdot\!\bar{c}$ & $c\!\cdot\!\bar{c}$ & $\bar{c}^2$ & $c^2$ & $c\!\cdot\!\bar{c}$ & $c\!\cdot\!\bar{c}$ & $\bar{c}^2$ & $\bar{j}^2$ & $j\!\cdot\!\bar{j}$ & $j\!\cdot\!\bar{j}$ & $j^2$ \\ 
$um$ & $c\!\cdot\!\bar{d}$ & $c\!\cdot\!d$ & $\bar{c}\!\cdot\!\bar{d}$ & $\bar{c}\!\cdot\!d$ & $c\!\cdot\!\bar{d}$ & $c\!\cdot\!d$ & $\bar{c}\!\cdot\!\bar{d}$ & $\bar{c}\!\cdot\!d$ & $\bar{j}\!\cdot\!\bar{d}$ & $\bar{j}\!\cdot\!d$ & $j\!\cdot\!\bar{d}$ & $j\!\cdot\!d$ \\ 
$uh$ & $c\!\cdot\!\bar{e}$ & $c\!\cdot\!e$ & $\bar{c}\!\cdot\!\bar{e}$ & $\bar{c}\!\cdot\!e$ & $c\!\cdot\!f$ & $c\!\cdot\!\bar{f}$ & $\bar{c}\!\cdot\!f$ & $\bar{c}\!\cdot\!\bar{f}$ & $\bar{j}\!\cdot\!\bar{e}$ & $\bar{j}\!\cdot\!e$ & $j\!\cdot\!\bar{e}$ & $j\!\cdot\!e$ \\ 
$uf$ & $c\!\cdot\!g$ & $c\!\cdot\!\bar{g}$ & $\bar{c}\!\cdot\!g$ & $\bar{c}\!\cdot\!\bar{g}$ & $c\!\cdot\!g$ & $c\!\cdot\!\bar{g}$ & $\bar{c}\!\cdot\!g$ & $\bar{c}\!\cdot\!\bar{g}$ & $\bar{j}\!\cdot\!g$ & $\bar{j}\!\cdot\!\bar{g}$ & $j\!\cdot\!g$ & $j\!\cdot\!\bar{g}$ \\ 
$mu$ & $c\!\cdot\!\bar{d}$ & $\bar{c}\!\cdot\!\bar{d}$ & $c\!\cdot\!d$ & $\bar{c}\!\cdot\!d$ & $c\!\cdot\!\bar{d}$ & $\bar{c}\!\cdot\!\bar{d}$ & $c\!\cdot\!d$ & $\bar{c}\!\cdot\!d$ & $\bar{j}\!\cdot\!\bar{d}$ & $j\!\cdot\!\bar{d}$ & $\bar{j}\!\cdot\!d$ & $j\!\cdot\!d$ \\ 
$mm$ & $\bar{d}^2$ & $d\!\cdot\!\bar{d}$ & $d\!\cdot\!\bar{d}$ & $d^2$ & $\bar{d}^2$ & $d\!\cdot\!\bar{d}$ & $d\!\cdot\!\bar{d}$ & $d^2$ & $\bar{d}^2$ & $d\!\cdot\!\bar{d}$ & $d\!\cdot\!\bar{d}$ & $d^2$ \\ 
$mh$ & $\bar{d}\!\cdot\!\bar{e}$ & $\bar{d}\!\cdot\!e$ & $d\!\cdot\!\bar{e}$ & $d\!\cdot\!e$ & $\bar{d}\!\cdot\!f$ & $\bar{d}\!\cdot\!\bar{f}$ & $d\!\cdot\!f$ & $d\!\cdot\!\bar{f}$ & $\bar{d}\!\cdot\!\bar{e}$ & $\bar{d}\!\cdot\!e$ & $d\!\cdot\!\bar{e}$ & $d\!\cdot\!e$ \\ 
$mf$ & $\bar{d}\!\cdot\!g$ & $\bar{d}\!\cdot\!\bar{g}$ & $d\!\cdot\!g$ & $d\!\cdot\!\bar{g}$ & $\bar{d}\!\cdot\!g$ & $\bar{d}\!\cdot\!\bar{g}$ & $d\!\cdot\!g$ & $d\!\cdot\!\bar{g}$ & $\bar{d}\!\cdot\!g$ & $\bar{d}\!\cdot\!\bar{g}$ & $d\!\cdot\!g$ & $d\!\cdot\!\bar{g}$ \\ 
$hu$ & $c\!\cdot\!\bar{e}$ & $\bar{c}\!\cdot\!\bar{e}$ & $c\!\cdot\!e$ & $\bar{c}\!\cdot\!e$ & $c\!\cdot\!f$ & $\bar{c}\!\cdot\!f$ & $c\!\cdot\!\bar{f}$ & $\bar{c}\!\cdot\!\bar{f}$ & $\bar{j}\!\cdot\!\bar{e}$ & $j\!\cdot\!\bar{e}$ & $\bar{j}\!\cdot\!e$ & $j\!\cdot\!e$ \\ 
$hm$ & $\bar{d}\!\cdot\!\bar{e}$ & $d\!\cdot\!\bar{e}$ & $\bar{d}\!\cdot\!e$ & $d\!\cdot\!e$ & $\bar{d}\!\cdot\!f$ & $d\!\cdot\!f$ & $\bar{d}\!\cdot\!\bar{f}$ & $d\!\cdot\!\bar{f}$ & $\bar{d}\!\cdot\!\bar{e}$ & $d\!\cdot\!\bar{e}$ & $\bar{d}\!\cdot\!e$ & $d\!\cdot\!e$ \\ 
$hh$ & $\bar{e}^2$ & $e\!\cdot\!\bar{e}$ & $e\!\cdot\!\bar{e}$ & $e^2$ & $f^2$ & $f\!\cdot\!\bar{f}$ & $f\!\cdot\!\bar{f}$ & $\bar{f}^2$ & $\bar{e}^2$ & $e\!\cdot\!\bar{e}$ & $e\!\cdot\!\bar{e}$ & $e^2$ \\ 
$hf$ & $\bar{e}\!\cdot\!g$ & $\bar{e}\!\cdot\!\bar{g}$ & $e\!\cdot\!g$ & $e\!\cdot\!\bar{g}$ & $f\!\cdot\!g$ & $f\!\cdot\!\bar{g}$ & $\bar{f}\!\cdot\!g$ & $\bar{f}\!\cdot\!\bar{g}$ & $\bar{e}\!\cdot\!g$ & $\bar{e}\!\cdot\!\bar{g}$ & $e\!\cdot\!g$ & $e\!\cdot\!\bar{g}$ \\ 
$fu$ & $c\!\cdot\!g$ & $\bar{c}\!\cdot\!g$ & $c\!\cdot\!\bar{g}$ & $\bar{c}\!\cdot\!\bar{g}$ & $c\!\cdot\!g$ & $\bar{c}\!\cdot\!g$ & $c\!\cdot\!\bar{g}$ & $\bar{c}\!\cdot\!\bar{g}$ & $\bar{j}\!\cdot\!g$ & $j\!\cdot\!g$ & $\bar{j}\!\cdot\!\bar{g}$ & $j\!\cdot\!\bar{g}$ \\ 
$fm$ & $\bar{d}\!\cdot\!g$ & $d\!\cdot\!g$ & $\bar{d}\!\cdot\!\bar{g}$ & $d\!\cdot\!\bar{g}$ & $\bar{d}\!\cdot\!g$ & $d\!\cdot\!g$ & $\bar{d}\!\cdot\!\bar{g}$ & $d\!\cdot\!\bar{g}$ & $\bar{d}\!\cdot\!g$ & $d\!\cdot\!g$ & $\bar{d}\!\cdot\!\bar{g}$ & $d\!\cdot\!\bar{g}$ \\ 
$fh$ & $\bar{e}\!\cdot\!g$ & $e\!\cdot\!g$ & $\bar{e}\!\cdot\!\bar{g}$ & $e\!\cdot\!\bar{g}$ & $f\!\cdot\!g$ & $\bar{f}\!\cdot\!g$ & $f\!\cdot\!\bar{g}$ & $\bar{f}\!\cdot\!\bar{g}$ & $\bar{e}\!\cdot\!g$ & $e\!\cdot\!g$ & $\bar{e}\!\cdot\!\bar{g}$ & $e\!\cdot\!\bar{g}$ \\ 
$ff$ & $g^2$ & $g\!\cdot\!\bar{g}$ & $g\!\cdot\!\bar{g}$ & $\bar{g}^2$ & $g^2$ & $g\!\cdot\!\bar{g}$ & $g\!\cdot\!\bar{g}$ & $\bar{g}^2$ & $g^2$ & $g\!\cdot\!\bar{g}$ & $g\!\cdot\!\bar{g}$ & $\bar{g}^2$ \\ 
\end{tabular}
\end{center}
\end{table}

\subsection{Parameter estimation}

Recall that the set of hidden states is $\mathcal{S}=\{uu, um , uh , uf , mu , mm , mh , mf , hu,$ $ hm , hh , hf , fu , fm , fh , ff \}$.
We now define a hidden Markov model (HMM) based on the 
model presented in Section 2.2.
As set of observable states we define $\mathcal{S}_{obs}=\{TT,TC,$ $CT, CC\}$, i.e. we use the results of the sequencing experiments (cf. Fig. \ref{Fig:ConvErr}) on both strands. The conversion errors define the corresponding emission probabilities.
We also define $n_e(j,t)$ as the number of times that state $j\in S_{obs}$ has been observed during independent measurements of sequencing method $e\in E:=\{\text{BS, oxBS, MAB-Seq}\}$.
The probability distribution over all observable states for experiment $e$ is denoted by $\pi_e(t)$, the probability of a state $j\in S_{obs}$ by $\pi_e(j,t)$ and in a similar fashion we denote $\pi(i,t)$ with $i \in S $ for the hidden states, with   probability distribution $\pi(t)$.
The observable and hidden states for all times $t$ are connected via
\begin{equation}
\pi_{e}(t)=\pi(t)\cdot E_{e},
\end{equation}
where $E_e$ is the emission matrix for sequencing method $e$ and is listed in Tab.~\ref{Tab:ConvErr} for each of the three methods.

Our goal is to estimate the efficiencies for the different methylation events given our hybrid HMM and data from the three different experiments at different time points $t\in T_{obs}$ via a maximum likelihood estimator (MLE).
Since an initial distribution over the hidden states, which can not directly be observed, is needed in order to initialize the model, we have to employ the MLE twice:
First we estimate the initial distribution $\pi(0)$ over the hidden states by maximizing
\begin{equation}
\pi(0)^*=\text{argmax}_{\pi(0)}\mathcal{L}_1(\pi(0)),
\end{equation}
under the constrain that $\sum_{i\in S} \pi(i,0)=1$. 
The likelihood $\mathcal{L}_1(\pi(0))$ is defined as
\begin{equation}
\mathcal{L}_1(\pi(0))=\prod_{e \in E} \prod_{j \in \mathcal{S}_{obs}} \pi_e(j,0)^{n_e(j,0)}.
\label{Eq:L1}
\end{equation}
Note that Eq.~\eqref{Eq:L1} is independent of the parameters.
Given an initial distribution over the hidden states we can now run our model and apply the MLE 
\begin{equation}
\vec{v}^*=\text{argmax}_{\vec{v}}\mathcal{L}_2(\vec{v}),
\end{equation}
a second time in order to estimate the efficiencies, where 
\begin{equation}
\mathcal{L}_2(\vec{v})=\prod_{e \in E} \prod_{t \in T_{obs}\backslash \{ 0 \} } \prod_{j \in \mathcal{S}_{obs}} \pi_e(j,t)^{n_e(j,t)}.
\end{equation}
The vector $\vec{v}=(\vec{v}_{\mu_m},\vec{v}_{\mu_d},\vec{v}_{\eta},\vec{v}_{\phi},\vec{v}_{\delta},p)$ contains all unknown parameters for all efficiencies and the probability $p$ of considering a hydroxylated cytosine as unmethylated.
Note that applying the MLE twice and independently leads only to an approximation of the true most likely explanation, since the estimated initial distribution may not lead to the same result in the parameter estimation, as if it would all be done in one estimation.
However, we choose this approach to reduce the computational complexity of the optimization.
In order to estimate the standard deviations of the estimated parameters we use the observed Fisher information matrix \cite{efron1978assessing}.
The Fisher information is defined as $\mathcal{J}(\vec{v}^*)=-\mathcal{H}(\vec{v}^*)$, where $\vec{v}^*$ is the maximum likelihood estimate and $\mathcal{H}(\vec{v})=\nabla\nabla^T\log \mathcal{L}_2(\vec{v})$ the Hessian matrix of the log-likelihood.
The expected Fisher information is then given by $\mathcal{I}(\vec{v}^*)=\mathbb{E}[\mathcal{J}(\vec{v}^*)]$ and its inverse forms a lower bound for the covariance matrix.
Thus, we can approximate the standard deviation of all estimated parameters by $\sigma(\vec{v}^*)=\sqrt{\text{Var}(\vec{v}^*)}\approx\sqrt{\text{diag}(-\mathcal{H}^{-1}(\vec{v}^*))}$.

The implementation of the hybrid HMM and its analysis as explained above has been integrated into the  latest beta version of the H(O)TA tool \cite{Kyriakopoulos2017}. H(O)TA provides results for individual CpGs and also an aggregated profile across all analyzed CpGs.

\section{Results}
In the following, we will discuss the results after applying our model to data 
derived from a short region at the single copy gene Afp (alpha fetoprotein), which contains 5 CpGs. 
More precisely, we followed the DNA methylation changes during the adjustment of mouse embryonic stem cells (mESCs) towards 2i medium after previous long time cultivation under Serum/LIF conditions \cite{ficz2013fgf,Giehr2016}. 
The Serum/ LIF-to-2i shift is a common model system which induces genome wide demethylation in mESCs including the Afp locus.
 
\begin{figure}[t]
\begin{center}
\includegraphics[scale=0.37]{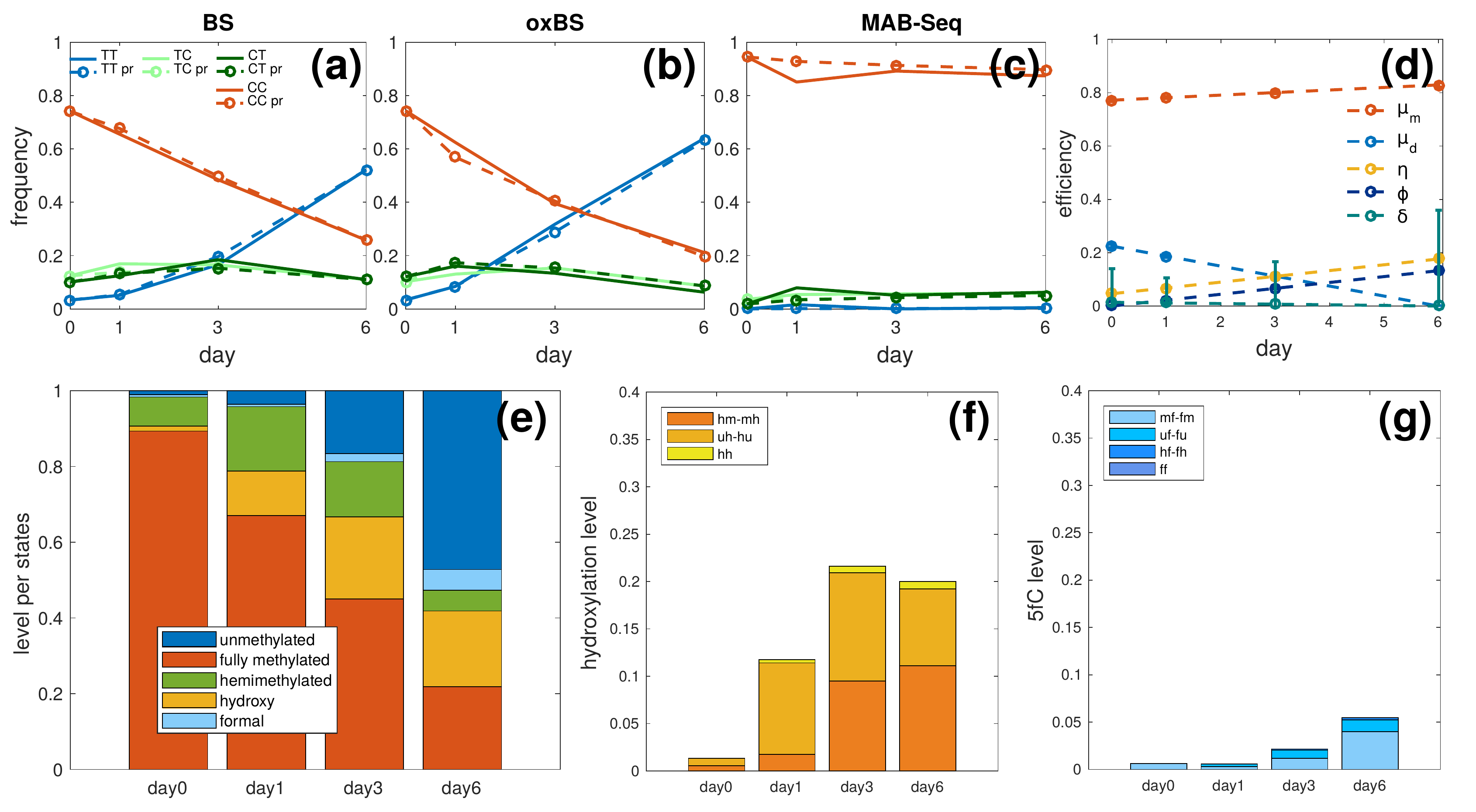}
\end{center}
\caption{Results for Afp. (a)-(c) Predicted frequencies (dashed lines) and frequencies obtained from sequencing experiments (solid lines) of the observable states for all three methods. (d) Estimated efficiencies with standard deviation. (e) Probabilities of the hidden states. (f) Detailed distribution of hydroxylated CpGs. (g) Detailed distribution of formalized CpGs.}
\label{Fig:Afp}
\end{figure}

Here, the individual and aggregated H(O)TA  results  show a very similar behavior. Hence, we only present the aggregated results  in Fig.~\ref{Fig:Afp}.

Fig.~\ref{Fig:Afp} (a) - (c) shows a comparison of the actual measurements (solid lines) and the results from the model (dashed lines) of the time evolution for the four observable states for all three sequencing models.
The predictions and measurements are in good agreement.
In BS and oxBS over time the frequency of TT increases, while the frequency of CC decreases.
Moreover, the frequencies of TC and CT 
show a temporal increase.
For 
MAB-Seq the frequencies of all observable states remain quite constant, except for some small changes in early times.
This behavior can be explained by the changes over time in the efficiencies, which are shown in Fig.~\ref{Fig:Afp} (d).
The maintenance probability $\mu_m$ remains constant at about $0.8$, as well as the probability $p$ of considering 5hmC as unmethylated, which is constant by definition.
The estimation of $p$ gives $p=1$, i.e. 5hmC is always considered to be unmethylated and will not be recognized by Dnmt1 after replication which means that 5hmC leads to an impairment of Dnmt1 activity and a passive loss of DNA methylation with each replication. 
The \emph{de novo} efficiency $\mu_d$ decreases over time, while the hydroxylation and 
formylation efficiency $\eta$ and $\phi$ increase.
The demethylation efficiency $\delta$ is always 0.
Note, however, that standard deviation becomes very large for later time points due to insufficient data.
The efficiencies explain the behavior of the frequencies of the observable states and is even more evident for the hidden states shown in Fig.~\ref{Fig:Afp} (e):
Over time the probability of being fully or hemimethylated decreases, while the probability of being unmethylated, hydroxylated or formylated increases.
A more detailed look into hydroxylated and formylated states is shown in Fig.~\ref{Fig:Afp} (f) and (g).
Note that the combination of hydroxylation and formylation is only shown in 
one of the two subplots, namely in (g).
The observed increase in 5hmC and 5fC/5caC is in accordance with the high oxidation efficiency of Tets in form of hydroxylation and formylation.
Previously, we showed using   a purely discrete HMM, that the presence of 5hmC leads to a block of Dnmt1 activity after replication, which we also observe in the present hybrid model \cite{Giehr2016}.
However, we now also observe an increase in higher oxidized cytosine variants, namely 5fC and 5caC which equally prevents methylation by Dnmt1.
Thus, we reason that the impact of Tet mediated oxidation of 5mC on DNA demethylation in the investigated system plays a much more important role than previously suggested \cite{von2016impairment}.

Considering the rather rapid periodic events of \emph{de novo} methylation and active demethylation, the chosen measurement time points are not ideal.
There is only information available at the end of each cell division cycle (one division within 24h), i.e., 
no information is given for the times between two cell divisions.
With measurements at time points between two cell divisions we would be able to distinguish if a CpG is in a certain state because it initially was, or because it ran through the full cycle, possibly multiple times. We would like to emphasize that with better data, i.e. with sufficiently many measurements between cell divisions, it will be possible to estimate all 
 reaction efficiencies with better confidence.

\section{Conclusion}
We proposed a hybrid hidden Markov model which is able to successfully describe both,  events  such as cell division and maintenance methylation that occur at fixed times, and events that occur at random times, 
such as \emph{de novo} methylation, oxidizations and active demethylation, according to a continuous-time Markov jump process.
To the best of our knowledge, this is the first model
that describes the dynamics of active demethylation, i.e., the active removal of the methyl group through 
several enzymatic steps.
We applied our model to  data from mouse embryonic stem cells, which undergo a gradually loss of DNA methylation over time.
We were able to accurately predict the frequency of the observable states and the levels of the hidden states in all cases.
We were also able to predict the enzymatic reaction efficiencies based on a linear assumption for their time behavior.

As future work  we plan to apply our model to more
informative data such that all efficiencies of the 
active demethylation cycle can be estimated with better confidence.
Moreover, we plan to allow different functional forms for the efficiencies since the linear form is  not flexible enough and does not allow to capture complex behaviors.
A suitable choice could be splines of different degrees and with a different number of knots.
However, in this case it is also necessary to perform model selection in order to prevent overfitting.
Another possible extension could be to investigate 
potential neighborhood dependencies of the modified cytosines \cite{luck2017stochastic}.

\bibliographystyle{splncs03}

\end{document}